\documentclass[10pt,conference]{IEEEtran}
\IEEEoverridecommandlockouts
\usepackage{booktabs}
\usepackage{cite}
\usepackage{amsmath,amssymb,amsfonts}
\usepackage{algorithmic}
\usepackage{graphicx}
\usepackage{textcomp}
\usepackage{bm}
\usepackage{xcolor}
\usepackage{graphicx}
\usepackage{url}
\usepackage{fancybox}
\usepackage{hyperref}
\usepackage{pifont}

\newcommand{\cmark}{\ding{51}}
\newcommand{\xmark}{\ding{55}}

\def\BibTeX{{\rm B\kern-.05em{\sc i\kern-.025em b}\kern-.08em
    T\kern-.1667em\lower.7ex\hbox{E}\kern-.125emX}}

\begin{document}

\title{From Conversation to Contribution: Characterizing Vibe Coding in Open-Source Software}

\author{
\IEEEauthorblockN{
Zihan Fang\IEEEauthorrefmark{1},
Yueke Zhang\IEEEauthorrefmark{1},
Ningzhi Tang\IEEEauthorrefmark{2},
Collin McMillan\IEEEauthorrefmark{2},
Toby Jia-Jun Li\IEEEauthorrefmark{2},
and Yu Huang\IEEEauthorrefmark{1}
}
\IEEEauthorblockA{
\IEEEauthorrefmark{1}Vanderbilt University, Nashville, TN, USA
}
\IEEEauthorblockA{
\IEEEauthorrefmark{2}University of Notre Dame, Notre Dame, IN, USA
}
}

\maketitle

\begin{abstract}
AI coding assistants such as GitHub Copilot and Cursor have evolved from code-suggestion tools into conversational collaborators, enabling \textit{vibe-coding} workflows in which developers guide AI-generated code through natural-language dialogue. 
Although researchers have increasingly recognized the importance of AI coding agents and begun examining their impact on open-source development, a comprehensive understanding of how developers’ chat-based interactions with AI relate to subsequent open-source development and collaboration remains limited.
This hinders efforts to effectively design, evaluate, and govern AI-assisted open-source software development.
To address this gap, we collected 13{,}360 AI conversation sessions comprising 79{,}172 user messages from 1{,}356 OSS repositories, linked them to repository development histories, and complemented this analysis with a targeted developer survey.
We find heavier AI use in smaller, less mature, and less collaborative repositories.
After AI adoption, projects tended to show more active contributors and lower contributor concentration ($p<.001$), although communication remained highly concentrated.
Code Writing was the dominant chat purpose, and nearly all AI chat sessions were followed by subsequent commits.
We find no broad deterioration in code-quality signals or pull request merging rates. 
However, developers perceive others’ AI-generated code as harder to maintain than their own ($p = .029$) and view AI as lowering barriers to OSS contribution. 
While most (68\%) are willing to share their chat, concerns remain around appearing incompetent, increasing reviewer burden, and exposing ideas to competitors.
These findings provide a large-scale empirical characterization of AI-assisted OSS contribution and offer practical insights for designing and governing responsible vibe-coding practices in open-source development.
\end{abstract}


\section{Introduction}
AI coding assistants such as GitHub Copilot and Cursor have become increasingly common in modern software development~\cite{chen2026code, solohubov2023accelerating}. 
Compared with general-purpose browser-based chat interfaces such as ChatGPT, these tools operate directly within developers’ programming environments and can access local project context. 
They support multi-turn dialogue, coordinated multi-file edits, terminal interaction, and iterative refinement based on codebase and execution feedback. 
Developers can therefore express high-level intent in natural language and steer AI-generated changes through dialogue, using the assistant as a conversational collaborator within the development environment.
This style of interaction is often described as \textit{vibe coding}: an intent-driven workflow in which developers express goals through natural-language prompts and rely on AI-generated code to varying degrees of manual oversight~\cite{chen2026programmers,ray2025review}.

Open-source software (OSS) provides a particularly relevant context for studying \textit{vibe coding} workflows because many AI-assisted development activities are observable in public repositories’ development histories, such as commits, issues, and pull requests (PRs)~\cite{kalliamvakou2014promises,lin2025open, li2025rise}. 
Understanding how developers make \textit{vibe-coding}-based OSS contributions is important because OSS is a collaborative ecosystem, where AI-assisted coding may affect not only individual productivity but also contribution practices, maintenance work, and community trust. Recent industry reports suggest that AI-assisted development is already straining GitHub’s infrastructure: code changes were reportedly on pace to increase from about 1 billion in 2025 to 14 billion in 2026, prompting Microsoft to explore additional multi-cloud capacity~\cite{stewart2026githubcapacity}.
These trends raise concerns about AI-assisted development in OSS: while it may improve efficiency, accelerate implementation, and revive inactive projects, it may also increase risks of insufficient review, maintenance burden, reduced testing, and less deliberate engineering decisions~\cite{peng2023impact, xu2025ai}.
Understanding these tradeoffs requires first characterizing how developers use AI coding assistants in OSS \textit{vibe-coding} workflows.
Although prior studies provide early evidence on AI adoption in OSS, they mainly examine observable outcomes such as commits, pull requests (PRs), and code-quality signals. For example, prior work reports that Cursor and autonomous-agent adoption can yield short-term velocity gains, but may also increase static-analysis warnings, code complexity, and post-merge churn~\cite{he2025speed, agarwal2026ai, popescu2026investigating}. Other studies find that agent-generated contributions are more likely to merge when supported by actionable reviewer engagement~\cite{robbes2026agentic,nachuma2026ai}. Less is known about the human--AI interaction process behind these outcomes, leaving open how \textit{vibe-coding} interactions shape subsequent OSS development activity.

Motivated by these gaps, we link developers’ conversational interactions with AI coding assistants to subsequent development activity, offering a more complete understanding of how vibe-coding workflows may influence OSS developers and communities.
Specifically, we collected 13{,}360 chat sessions between developers and AI coding assistants (e.g., Copilot, Cursor, and Claude Code) in OSS contributions. 
We examine how AI coding assistants are used in repositories, how repository dynamics change after adoption, and how developers whose chat histories appear in our dataset perceive and practice vibe coding in OSS.
We find that smaller, less mature, and less collaborative repositories tend to exhibit heavier AI use than their counterparts. After AI adoption, projects tended to show increases in active contributors and lower contributor concentration ($p<.001$), but communication remained concentrated among a few participants.
Furthermore, Code Writing was the dominant chat purpose, and nearly all AI chat sessions were followed by subsequent commits. 
We find no broad deterioration in observable code-quality signals, but survey responses show that developers were more concerned about others’ AI-generated code than their own and perceived others’ AI-generated code as imposing a greater maintenance burden ($p=.029$).
Respondents generally viewed AI as lowering barriers to OSS contribution and were often willing to share chat histories, but raised concerns about reputation, reviewer burden, and idea exposure.
We claim the following contributions:
\begin{itemize}
    \item A large-scale empirical characterization of vibe-coding-based AI coding-assistant adoption in OSS.
    \item A comprehensive understanding of how OSS dynamics change after AI adoption.
    \item A mixed-methods study linking vibe-coding behavior to subsequent OSS development activity and developer perceptions.
\end{itemize}

In this paper, we aim to provide practical insights for understanding AI's impact on OSS communities, designing more effective and responsible AI-assisted development practices, and informing future AI coding tool design.

\section{Related Work}
% We discuss prior work on how AI coding assistants have been developed and used in software development and OSS.
\subsection{AI Coding Assistants in Software Engineering}
Large language models (LLMs) have rapidly reshaped software engineering by enabling AI-assisted coding tools for code generation, completion, debugging, documentation, testing, and refactoring~\cite{fan2023large}. A growing body of software engineering research has focused on improving the techniques behind these tools. For example, prior work has explored feedback loops that iteratively refine generated code using execution results or test outcomes, as well as reinforcement learning methods that optimize code for efficiency, resource use, and developer comprehension~\cite{jiang2026survey, wang2024enhancing, fang2025dpo, zhang2025codeact}. More recently, LLM-based coding agents have extended these capabilities beyond isolated code generation, enabling more autonomous support for tasks such as task decomposition, debugging, and tool integration across the development lifecycle~\cite{liu2024large}. As a result, AI coding tools are increasingly framed as AI pair programmers that reduce implementation effort and accelerate routine development tasks~\cite{ajiga2024enhancing, abbas2025enhancing}.
Empirical studies show that AI coding tools can improve productivity and reduce development time. For example, Copilot users completed programming tasks faster~\cite{peng2023impact} and reported productivity gains reflected in behavioral usage data~\cite{ziegler2024measuring}.

At the same time, prior work highlights important limitations of AI-assisted coding. Studies of Copilot’s code suggestions show that generated solutions vary in correctness, complexity, robustness, and reproducibility~\cite{nguyen2022empirical,dakhel2023github}. Security-focused work further shows that Copilot may generate insecure code in security-sensitive scenarios~\cite{pearce2025asleep}. Human-centered studies add that, although developers often value Copilot’s suggestions, these tools do not always improve task completion and may introduce challenges in understanding, validating, and integrating generated code~\cite{vaithilingam2022expectation, tang2026coding}. 
Together, prior work shows that AI-assisted coding affects not only productivity but also how developers evaluate and maintain code. However, most studies focus on controlled tasks or short-term outcomes, leaving less known about how AI-assisted coding unfolds from developer--AI conversations to sustained activity in real-world repositories.

\subsection{AI Coding Assistant in Open Source Software}
Open-source software (OSS) provides an important setting for studying AI-assisted development because contributions are collaborative, publicly visible, and shaped by concerns about review, trust, reputation, and maintainability. It also offers an observable context for examining how developers use AI coding assistants in \textit{vibe-coding} workflows. 
Recent work has begun to examine generative AI in OSS contexts. Song et al. studied the impact of GitHub Copilot on collaborative OSS development and found evidence that Copilot use can increase project-level productivity and developer participation, but may also increase integration time, suggesting additional coordination costs~\cite{song2024impact}. 
Other work has examined AI support for pull-request workflows and studied generative AI for PR descriptions, showing how LLMs can assist with summarizing and documenting code changes in review contexts \cite{xiao2024generative}. 

Recent studies have shifted from controlled evaluations of AI coding tools to large-scale analyses of their use in OSS. Li et al. introduced AIDev, a dataset of agent-authored pull requests from multiple autonomous coding agents, and showed that agents are beginning to act as software contributors by opening PRs, participating in review workflows, and producing code changes at scale~\cite{li2025rise}. Complementary causal studies examine repository-level effects after tool adoption: He et al. found that Cursor adoption produces short-term increases in development velocity but also persistent increases in static-analysis warnings and code complexity~\cite{he2025speed}, while Agarwal et al. showed that autonomous agents yield front-loaded velocity gains mainly in repositories without prior AI-tool use, alongside sustained increases in warnings and cognitive complexity~\cite{agarwal2026ai}. Other work analyzes agent-generated artifacts more directly: Robbes et al. found that agent-assisted commits are larger than human-only commits and frequently involve feature development or bug fixing~\cite{robbes2026agentic}, Popescu et al. linked autonomous-agent contributions to greater post-merge churn~\cite{popescu2026investigating}, and Nachuma et al. showed that agent-authored PRs are more likely to be integrated when they receive actionable reviewer engagement, whereas larger or coordination-disrupting changes are less likely to be merged~\cite{nachuma2026ai}. These studies provide important evidence on agent-authored commits and pull requests, but they leave open how developers’ IDE-based conversations with AI coding assistants connect to subsequent repository activity. They also provide limited insight into how OSS developers who actively use these tools perceive their effects on participation and project development. Our work focuses on these gaps.

\section{Methodology}
We collected 13{,}360 AI chat sessions from 1{,}356 OSS repositories and complemented them with each repository's full development history.
Using these data, we investigated the following research questions:
\begin{itemize}
    \item \textbf{RQ1:} How do OSS repositories use AI coding assistants?
    \item \textbf{RQ2:} How do OSS repositories' dynamics change after AI adoption?
    \item \textbf{RQ3:} How do developers perceive AI tool usage through vibe coding in OSS?
\end{itemize}

\begin{figure}[t]
    \centering
    \includegraphics[width=0.8\linewidth]{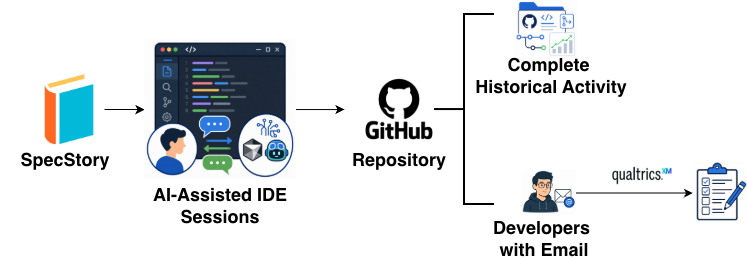}
    \caption{Data collection pipeline}
    \label{fig:data-collect}
\end{figure}

\subsection{Data Collection}
\label{sec:data-collection}
Figure~\ref{fig:data-collect} shows the data-collection pipeline. Our \href{https://anonymous.4open.science/r/vibe_coding_oss-CFE4/}{replication package} includes all scripts and materials.
\subsubsection{AI Chat Sessions}
Following prior work~\cite{tang2026programming}, we collected developers' AI-coding-assistant conversations from real-world conversational programming workflows using SpecStory, which preserves timestamped Markdown chat logs under \textit{.specstory/history/} for tools such as Cursor, GitHub Copilot, and Claude Code\footnote{\url{https://specstory.com/}}. We located these logs with GitHub Code Search and downloaded the raw Markdown files through the Git Blobs API, and parsed conversations using SpecStory's role
markers.
Each AI-chat session can contain multiple user messages. Following prior data-collection practices, we collected 13,360 AI-chat sessions containing 79,172 user messages, spanning September 2024 to March 2026, from 1,356 repositories.

\subsubsection{GitHub Repository Data}
To contextualize these chat sessions within broader development activity, we augmented the dataset with the full history of each GitHub repository. 
We used the GitHub REST API to collect complete repository histories, including creation and update timestamps, size, primary language, stars, forks, owner type, default branch, fork/archive status, contributors, commits, file-level changes, PRs, issues, comments, reviews, and CI/check-run records.
Specifically, for commit details, we first queried the commit-listing endpoint, \textit{GET /repos/{owner}/{repo}/commits}, to collect API-visible commits. Because this endpoint is branch-dependent, we also enumerated repository branches and tags using \textit{GET /repos/{owner}/{repo}/branches} and \textit{GET /repos/{owner}/{repo}/tags}, then queried commits from each branch and tag reference and de-duplicated commits by SHA. For each unique SHA, we requested detailed commit metadata using \textit{GET /repos/{owner}/{repo}/commits/{ref}}, which returns the commit timestamp, SHA, message, author and committer information, parent count, and file-level diff metadata, including changed file paths, file status, additions, deletions, and total changes.
We then applied a rule-based path classifier to categorize changed files as source code, tests, documentation, dependencies, configuration/build files, AI-chat artifacts, generated or binary assets, or other. 
Because file-type classification from paths is heuristic, we manually inspected a random sample of 100 classified files, identifying an initial misclassification rate of 3.0\%, and refined the rules accordingly to reduce error.

Of the 1,356 repositories, 1,287 had complete and
accessible histories; repositories that were deleted, private, or otherwise inaccessible were excluded.
To reduce noise from trivial projects, we followed prior work on identifying valid repositories~\cite{pickerill2020phantom} and screened our sample using signals of substantive development (e.g., development history, community activity, repository naming conventions, and README content). 
This process excluded 47 repositories (3.7\%) judged to be toy, homework, template, or otherwise trivial projects. Contributor identities were resolved using GitHub login, where available, and hashed email addresses otherwise; recognized bot and automated-agent accounts were excluded from all human-contributor analyses.
The final dataset comprised 1,240 repositories created between April 2013 and March 2026, with 12,108 AI-chat sessions retained for analysis and a total of 2,369 contributors across the dataset. The most common primary languages were TypeScript (25.8\%), Python (22.8\%), and JavaScript (12.9\%). In total, we collected 657,971 commit records, 1,400,804 changed-file records, 12,747 issues, 9,510 pull requests, 25,418 issue or pull-request comments, 6,902 pull-request reviews, and 120,489 CI/check records.
% To avoid rate-limit truncation, the collector used multiple API
% tokens, switched tokens when necessary, and waited for reset times if all tokens
% were exhausted. 
% Collection status was tracked by repository and data type,
% allowing failed or incomplete collections to be resumed. In the final collection
% log, no repositories were excluded because of unresolved rate limits; incomplete
% records were due to repository inaccessibility or API errors.

\subsubsection{Chat-Purpose and Project-Type Labels}
We labeled chat purposes using the behavioral-intent taxonomy and
LLM-based classification procedure from prior work~\cite{tang2026programming}.
This multi-label taxonomy comprises seven categories: Code Writing
(writing or editing code), Failure Reporting (reporting errors),
Delegation (asking the AI to perform tasks), Inquiry (asking for
explanations), Context Specification (providing constraints or
background), Workflow Control (turn-taking cues, e.g., ``continue" or confirmations), and Validation (checking correctness, quality, tests, or security). The GPT-5 classifier achieved a macro-F1 of 0.83 against a human-annotated reference set.
The resulting distribution, summarized in Table~\ref{tab:chat-purpose-distribution}, is broadly consistent with prior observations~\cite{tang2026programming}.
% The most common labels were Code Writing (34.5\%), Failure Reporting (24.0\%), and Inquiry (19.2\%), 
% Delegation (16.5\%), Context Specification (14.1\%), Workflow Control (11.5\%), and Validation (4.0\%).
Furthermore, we labeled project type using the taxonomy and LLM-plus-manual procedure from~\cite{fang2026contribution}. Because the taxonomy is multi-label, each repository could be assigned to one or more categories: Technology (90.2\%), Management-Productivity (33.1\%), and Creativity-Entertainment (5.6\%). In addition, 9.3\% of repositories were classified as OSS4SG. Manual verification on a random sample of 200 projects yielded an accuracy of 90.5\%.
% \paragraph{Project type.}
% Project content and social-good purpose were inferred from repository names,
% descriptions, topics, and available README text using classifiers trained on
% 403 manually labeled reference projects from the prior OSS/OSS4SG taxonomy.
% Content was modeled as a multi-label problem over 12 topics, with a probability
% threshold of 0.40 selected from cross-validated threshold diagnostics.
% Repositories could therefore receive zero, one, or multiple content labels.
% OSS4SG predictions were treated as screening candidates because the reference
% method calls for manual verification of positive predictions. These inferred
% labels were used only for exploratory association analyses and were not treated
% as ground-truth project categories.

\subsubsection{Survey Data}
We complemented the repository-activity analysis with a developer survey on perceptions of vibe-coding-based AI use in OSS development. Our institutional IRB approved the study. From 1,240 repositories, we identified 652 developers with publicly available email addresses and, after removing bounced emails, delivered 589 survey invitations via Qualtrics\footnote{\url{https://www.qualtrics.com/}}. Upon completion, participants were entered into a lottery for a \$30 Amazon gift card, with one gift card awarded per 20 participants. We received 25 responses, yielding a 4.2\% response rate among delivered invitations, consistent with response rates reported in prior software-engineering surveys~\cite{acar2017security, liang2022understanding}.

The consent screen required participants to confirm that they were at least 18 years old, had programming or OSS experience, and voluntarily agreed to participate. The survey measured participants' perceived benefits and risks of AI coding assistants in vibe coding; perceptions of code quality and maintainability; social and reputational concerns; authorship and responsibility; willingness to disclose AI use or share chat history; appropriate use cases; and open-ended reflections on benefits and risks. The survey included Likert-scale, multiple-choice, and open-ended questions and was designed to be completed within 15 minutes.
Most respondents are software developers (36\%) or students (20\%). They were generally experienced programmers: 92\% reported at least intermediate experience, and 72\% had programmed for three or more years. They were also familiar with vibe coding, with 76\% reporting very or extreme familiarity and 84\% using it for OSS contributions.

% The survey included 32 Likert-scale questions, two multiple-choice questions, and six open-ended questions.

\subsection{Data Analysis}
\label{sec:data-analysis}
\subsubsection{Preprocessing and temporal alignment}
\label{sec:data-analysis-0}
Because developers may use AI coding assistants during local development before publishing a repository to GitHub, SpecStory chat histories can predate the repository's GitHub \texttt{created\_at} timestamp, which reflects when the repository was created on GitHub, not when local development began.
We therefore define a repository's \textit{first observed AI adoption} as the timestamp of its earliest AI-chat session in our dataset. We mapped chat sessions to repositories using the session SHA and the GitHub location where the chat-history file was found. For each repository, we selected the earliest mapped chat session and converted all timestamps to UTC. Under this definition, 632 of the 1{,}240 repositories had their first observed AI chat at or before the GitHub repository creation timestamp, while the remaining 608 had their first observed AI chat after GitHub repository creation.

\subsubsection{RQ1: How do OSS repositories use AI coding assistants}
\label{sec:data-analysis-1}
For each chat-history file identified through GitHub Code Search, we used the \textit{commit\_or\_ref} in the GitHub blob URL to identify the repository state in which the artifact was observed. 
When the ref matched a commit in our collected history, we treated that commit as the artifact-containing commit for the session. 
We further classified these commits as \textit{mixed AI-artifact development commits} if they also modified development files, and as \textit{chat-history-only artifact commits} otherwise. In the following analyses, \textit{AI-related commits} refers to all artifact-containing commits, while analyses of AI-related development activity focus on mixed AI-artifact development commits.\\
% when multiple sessions linked to the same commit, the commit was assigned to the closest preceding session to avoid double-counting.\\
\textbf{RQ1-1: How are AI coding assistants used in OSS contribution?}
% For each repository $i$, AI adoption was defined as the first mapped AI-chat timestamp $T_i$, and analysis was restricted to post-adoption activity ($t \geq T_i$). 
% A commit was classified as AI-related if it changed AI chat history or artifact files, or combined source-code changes with AI-related artifacts. 
For each repository, we computed the share of AI-related commits among all post-adoption commits. 
Calendar months were indexed relative to the adoption month, with the adoption month coded as relative month 0 and excluded from pre/post comparisons to avoid transition-period effects.
We then assessed its temporal trend using Spearman correlation across relative months and a binomial regression modeling each post-adoption commit as AI-related or not, with relative month as the predictor.
% To characterize the scope, we excluded AI-artifact files and analyzed development-file changes only. 
For each AI-related commit, we computed the number of files changed, file-type categories touched, and total churn, measured as additions plus deletions, after excluding AI-artifact files.
% we separately computed source-code churn to test whether changes in total churn reflected source-code edits or other file types. 
% We modeled log-churn as a function of relative month using linear regression to assess temporal trends.
To assess associations with repository characteristics, we used Spearman correlations for numeric predictors, given the skewness of both predictors and outcomes, and then fit a multivariable binomial regression with AI-related commit count as the outcome and repository and contributor characteristics as predictors.
Finally, to assess clustering, we also excluded AI-artifact files and computed concentration at the file and module (top-level directory) levels: the share of AI-related file changes attributed to the single most-changed file or module, the share attributed to the top three, and the Herfindahl--Hirschman Index (HHI),

{\small
\[
\small
HHI = \sum_j s_j^2,
\]
}
where $s_j$ is the share of AI-related file changes attributed to file or module $j$. Higher HHI values indicate greater concentration, consistent with its use as a measure of inequality in software-maintenance effort~\cite{narayanan2009matter}.\\
\textbf{RQ1–2: How do developers chat with AI in OSS contributions?} We next examined how chat purpose is associated with subsequent development activity. 
We computed the average chat sessions per repository, purposes per repository, and dominant-purpose distributions, defining the dominant purpose as the category with the most labeled sessions in a repository.
% At the session level, we identified subsequent commits in the same repository. 
We then measured whether each chat session was followed by any commit or source-code commit. 
For the subsequent development commit, we classified the file types touched and grouped rare combinations as ``Less common mixes.'' For commit-scope analyses, we measured total and source-code churn, the number of source files changed, and the number of file categories touched.
Finally, we compared dominant chat-purpose distributions across project content categories and OSS4SG status.

\subsubsection{RQ2: How do OSS repositories' dynamics change after AI adoption}
As discussed in Section~\ref{sec:data-analysis-0}, we excluded repositories with pre-publication AI chats, since their observed AI use began before GitHub publication. 
Of the 1,240 repositories, 608 (49.0\%) had their first observed AI chat after GitHub publication; we use this group as the main sample. Because these repositories tended to be young, we additionally validated our results on a more mature and larger cohort of 114 repositories filtered from the main group, averaging 5.0 contributors and 3.5 years of age.
% 892.8 days of pre-adoption history, and 295.4 days of post-adoption follow-up. 
We compute the following measures for both the main cohort ($n=608$) and the validation cohort ($n=114$). 
To handle imbalanced pre- and post-adoption periods, we used normalized measures where appropriate and modeled temporal trends with the relative-month ITS specification.\\
\textbf{RQ2-1: How does repository commit activity change?} 
We analyzed repository activity at the repository-month level using the adoption timestamp defined in Section~\ref{sec:data-analysis-0}. For each relative month, as defined in Section~\ref{sec:data-analysis-1}, we computed average total commits and source-code commits before and after adoption.
To evaluate temporal trends before and after AI adoption, we estimated an interrupted time-series (ITS) model for each outcome:

{\small
\[
\begin{aligned}
\log(1+Y_{im}) ={}& \alpha_i + \beta_1 \text{PreTime}_{im}
+ \beta_2 \text{Post}_{im} \\
&+ \beta_3 \text{PostTime}_{im}
+ \beta_4 \log(1+\text{Age}_{im}) + \epsilon_{im}.
\end{aligned}
\]
}

where \(Y_{im}\) is the outcome for repository \(i\) in relative month \(m\), and \(\alpha_i\) is a repository fixed effect. \(\text{PreTime}_{im}\) estimates the pre-adoption trend, \(\text{Post}_{im}\) estimates the immediate level change after adoption, and \(\text{PostTime}_{im}\) estimates the post-adoption trend. Repository fixed effects control for stable repository differences, while \(\text{Age}_{im}\) adjusts for repository aging over time. Standard errors were clustered by repository, and we compared pre- and post-adoption trajectories using the slope contrast \(\beta_3-\beta_1\).
We use the same relative-month ITS specification for all later temporal trend analyses.\\
\textbf{RQ2-2: Is AI adoption associated with changes in code-quality signals in OSS repositories?} We evaluated code-quality signals using defect-related activity, testing behavior, and CI outcomes, following prior repository-mining practice~\cite{sliwerski2005changes, beller2017oops}. All measures were computed per repository for the pre- and post-adoption periods and then modeled using the relative-month ITS specification described above.

First, we measured defect-related activity using the number and share of bug/fix commits and bug-labeled issues. Bug/fix commits were identified from commit messages containing corrective terms such as \textit{fix}, \textit{bug}, \textit{error}, \textit{issue}, \textit{patch}, and \textit{regression}. 
Bug-labeled issues were identified from GitHub issue labels containing terms such as \textit{bug}, \textit{defect}, \textit{regression}, and \textit{error}.
Second, we measured testing and CI signals using the share of commits touching test files, CI failure rate, and CI success rate. 
A commit was considered test-touching if it modified at least one test file, identified using path- and filename-based rules such as \textit{test}, \textit{tests}, \textit{spec}, \textit{\_\_tests\_\_}, \textit{unittest}, \textit{pytest}, \textit{jest}, and \textit{mocha}. CI outcomes were collected from GitHub check-run and check-suite records: completed checks with \textit{success} were treated as successful, while \textit{failure}, \textit{timed\_out}, \textit{cancelled}, and \textit{action\_required} were treated as failed. Checks with missing, non-completed, or other terminal conclusions were excluded.\\
% Test-line churn was computed as additions plus deletions in files classified as test files. 
% \\
% Third, we measured downstream maintenance signals at the repository level using the share of eligible source commits followed by same-file rework within 1, 7, and 30 days; the share of eligible source commits followed by same-file bug/fix follow-up within 1, 7, and 30 days; and the share of revert commits. 
% Finally, among commits that received downstream follow-up, we computed repository-level time interval to same-file rework and same-file bug/fix follow-up.
\textbf{RQ2-3: How do Issues and Pull Requests change?} We analyzed issue and pull-request activity by volume, share, and resolution/merge efficiency. Using the first observed AI-chat timestamp as the cutoff, we assigned openings by creation time, issue closings by closure time, and merged PRs by merge time, then compared pre/post counts per repository. For share-based measures, we computed issue and PR shares among opened items, issue closure-to-opening ratios, and PR merge shares. For efficiency, issues and PRs were assigned by creation time; issue-resolution time measured creation to closure, and PR-merge time measured creation to merge. We computed mean and median times per repository before and after adoption, including only repositories with at least one resolved issue or merged PR in both periods.\\
\textbf{RQ2-4: Is AI adoption associated with changes in OSS collaboration patterns?} We defined active contributors as non-bot contributors with at least one commit in a repository period. Contributor participation was measured by the number of active contributors, and concentration by the top-contributor share,

{\small
\[
\text{TopContributorShare}_{rp} =
\frac{\max_j(\text{commits}_{jrp})}{\sum_j \text{commits}_{jrp}},
\]
}

where $\text{commits}_{jrp}$ is the number of substantive commits by contributor $j$ in repository $r$ during period $p$, and by the HHI defined in Section~\ref{sec:data-analysis-1}.
For communication and review activity, we assigned issues, pull requests, issue/PR comments, PR review comments, and PR reviews to periods by creation timestamp. We measured intensity using comments per issue/PR, PR review comments per PR, unique issue/PR commenters, and unique PR reviewers. We measured concentration using the top-commenter share, defined analogously to the top-contributor share above but over issue/PR comments, and the commenter and reviewer HHIs (see Section~\ref{sec:data-analysis-1}).
% All measures above were computed separately for the main sample and the older validation cohort described in Section~\ref{sec:validation-cohort}.

\subsubsection{RQ3: How do developers perceive AI tool usage through vibe coding in OSS}
We analyzed survey responses using descriptive and construct-level summaries. Five-point Likert items were coded from 1 (strongly disagree) to 5 (strongly agree) and summarized using means and the percentage of respondents selecting somewhat or strongly agree. For multi-item constructs, such as OSS contribution access, we averaged related items per respondent and summarized construct scores using the mean and percentage above the neutral midpoint. We used within-respondent comparisons for paired perception items, such as concerns about one's own versus others' AI-generated code. 
Multiple-choice questions were summarized as the proportion of valid respondents selecting each option, and open-ended responses were reviewed to contextualize quantitative patterns.

\subsubsection{Statistical inference.}
We selected statistical tests based on the structure of each analysis. For repository-level before--after comparisons, we computed one pre-adoption and one post-adoption value per repository and used paired Wilcoxon signed-rank tests, as many measures were sparse and non-normally distributed. For categorical associations, such as relationships between dominant chat purpose and project characteristics, we used chi-square tests or permutation tests where appropriate. For skewed continuous outcomes compared across chat categories, we used non-parametric tests; when multiple chat-linked observations came from the same repository, we used repository-blocked permutation tests.
We corrected for multiple comparisons using the Benjamini--Hochberg false-discovery-rate procedure within families of related tests rather than globally across all analyses. Throughout, we report raw $p$-values, FDR-adjusted $q$-values, and model coefficients or odds ratios where applicable.
\begin{table}[t]
\centering
\scriptsize
\setlength{\tabcolsep}{4pt}
\renewcommand{\arraystretch}{1.02}
\caption{Chat-purpose distribution and subsequent commit churn. The dominant repository purpose is the most frequent chat label per repository. Churn reports mean/median lines changed.}
\label{tab:chat-purpose-distribution}
\begin{tabular}{@{}lcccc@{}}
\toprule
\textbf{Purpose} & 
\textbf{Sess.} & 
\textbf{Dom. repo.} & 
\textbf{Src. churn} & 
\textbf{Total churn} \\
 & 
\textbf{share} & 
\textbf{share} & 
\textbf{mean/med.} & 
\textbf{mean/med.} \\
\midrule
Code Writing        & 34.7\% & 53.9\% & 3{,}792 / 88  & 97{,}755 / 1{,}224 \\
Failure Reporting   & 24.2\% & 12.7\% & 2{,}797 / 32  & 16{,}400 / 293 \\
Delegation          & 16.0\% & 15.5\% & 3{,}342 / 37  & 18{,}460 / 1{,}104 \\
Inquiry             & 19.3\% & 8.9\%  & 2{,}152 / 7   & 25{,}246 / 205 \\
Context Spec.       & 14.1\% & 5.3\%  & 2{,}574 / 156 & 18{,}931 / 958 \\
Workflow Control    & 11.5\% & 3.3\%  & 1{,}982 / 115 & 10{,}213 / 644 \\
Validation          & 3.9\%  & 0.4\%  & 1{,}728 / 0   & 15{,}835 / 48 \\
\bottomrule
\end{tabular}
\vspace{-8pt}
\end{table}

\section{Results}
\subsection{RQ1: How do OSS repositories use AI coding assistants?}
\label{sec:result-1}

\subsubsection{How are AI coding assistants used in OSS contributions}
% We characterized AI-related commit activity following adoption in terms of its volume, scope, and clustering.\\
\textbf{AI-related commit prevalence and size.} 
Across the 1,240 repositories, AI-related commits accounted for 44.3\% of post-adoption commits on average (median: 34.3\%).
AI-related commits share declined sharply over time, from 32.8\% of post-adoption
commits in the first month to 16.1\% by six months and 4.4\% by twelve
months. This downward trend was confirmed by a Spearman correlation ($\rho = -0.940$, $p < .001$) and a commit-level binomial regression (OR $= 0.762$ per month, $p = .015$), indicating that AI assistance was
used most heavily right after adoption and progressively phased out as repositories matured.

Moreover, across AI-related commits, total churn was highly right-skewed, with a mean of 19{,}947 lines and a median of 1{,}789. 
Although total churn fluctuated over time, a log-scale model estimated a 7.1\% increase per relative month after adoption ($p=.007$). 
In contrast, source-code churn decreased by 13.2\% per month ($p=.002$). 
Mean churn declined from 2{,}211 lines in the adoption month to 1{,}456 in months 4--6 and 1{,}911 in months 7--12; medians fell from 113 to 28.5 and 32 lines, respectively. 
Thus, the increase in total churn appears to be driven by non-source changes or unusually large commits, rather than larger source-code edits.\\
% commit churn - from 15,473 lines (median = 1,639) in the adoption month to 24,485 (median=2,024) in month 1, 19,188 in month 6 (median = 4,015). 
\textbf{Repository characteristics and AI-use intensity.}
We next examined whether repository characteristics were associated with AI-use intensity. In bivariate analyses, AI-related commit share was negatively correlated with repository size, age, contributor count, stars, forks, and total commits, with the strongest association observed for post-adoption commits ($\rho=-0.570$, $p<.001$). This association partly reflects a denominator artifact, since AI-related commit share is calculated using total commits as the denominator, but it also indicates that AI-related commits did not increase proportionally with overall repository activity.
After adjusting for repository characteristics in a multivariable binomial model, only repository size in MB (OR $=0.834$, $p=.010$) and contributor count (OR $=0.531$, $p=.047$) remained significant negative predictors. Repository age, programming language, and project domain showed no robust associations.
Overall, AI-related commits were more prominent in repositories with smaller codebases and fewer contributors, even after accounting for repository age, language, and domain.\\
\textbf{Scope and concentration of AI-related changes.} 
We also examined the scope of files touched by AI coding assistants. 
On average, each AI-related commit changed 25.7 files (median = 7.0) across 3.3 file-type categories (median = 3). 
Source-code files made up the largest share of files changed (43.6\%), followed by documentation (15.0\%), generated or binary files (11.8\%), configuration files (3.5\%), test files (3.2\%), and dependency files (2.0\%). 
The remaining 20.9\% consisted of miscellaneous or unclassified files not assigned to a distinct category. 

We further examined whether AI-related file changes were concentrated in specific files or modules (a module being the top-level directory in a file path, e.g., \texttt{src}, \texttt{tests}, with root-level files grouped as \texttt{(root)}). Concentration was low at the file level: the top file accounted for 10.7\% of AI-related changes on average (median = 5.0\%), the top three for 23.6\% (median = 13.0\%), and HHI averaged 0.08 (median = 0.03). Concentration was higher at the module level: the top module accounted for 62.2\% on average (median = 59.9\%), the top three for 90.8\% (median = 97.4\%), and HHI averaged 0.52 (median = 0.46).
Thus, AI-related work spanned many files but clustered within a few repository areas, partly reflecting coarser module-level aggregation.

\subsubsection{How do developers chat with AI in OSS contributions}
% As described in Section~\ref{sec:data-analysis}, we classified developers' AI chats into seven purpose categories. We next examined how these chat purposes associated with subsequent commits.\\
\textbf{Distribution of chat purposes.} 
Repositories had an average of 9.8 chat sessions (median = 2.0) and used an average of 2.1 distinct chat purposes (median = 2.0). 
For each repository, we defined the dominant chat purpose as the category with the largest number of labeled sessions. Under this definition, Code Writing was dominant in 53.9\% of repositories, followed by Delegation (15.5\%), Failure Reporting (12.7\%), and Inquiry (8.9\%), as summarized in Table~\ref{tab:chat-purpose-distribution}.\\
% Context Specification (5.3\%), Workflow Control (3.3\%), and Validation (0.4\%).\\
\textbf{Subsequent activity by chat purpose.}
We next examined development activity in the commit associated with each
labeled chat session, referred to as subsequent activity. 
Nearly all
sessions (98.9\%) had subsequent activity, and 96.1\% were associated with
commits that changed at least one development file. 
We then characterized the file-type composition of these commits using the multi-label chat-purpose assignments. 
Across all purpose categories, the most common associated commit type was a source-only commit, accounting for 26.9\%--30.4\% of associated development commits depending on the purpose. 
Documentation-only commits were generally the second most common pattern, ranging from 14.8\% to 20.2\%. 
Workflow Control and Validation had the highest shares of broad commits touching source, tests, documentation, configuration, and dependency files together (10.3\% and 10.4\%, respectively). 
After accounting for repository-level clustering (i.e., that
observations from the same repository are not independent) using a
repository-blocked permutation test, we found no evidence that chat
purpose statistically predicted the file-type composition of the
associated commit.\\
\textbf{Subsequent activity scope by chat purpose.} 
We then examined the size and scope of the first subsequent development commit associated with each chat session, regardless of file type. 
As shown in Table~\ref{tab:chat-purpose-distribution}, total churn varied descriptively across chat purposes, with Code Writing showing the largest mean total churn and Validation the smallest. However, repository-blocked permutation tests found no robust differences in total churn, total files changed, or the number of file-type categories touched. 
Thus, the chat purpose did not statistically predict larger or broader subsequent commits overall.
When focusing only on source-code changes, source-code churn differed significantly by chat category (\(p=.001, q=.007\)), with Failure Reporting, Delegation, and Code Writing showing the largest mean source churn. 
However, the number of source files changed did not differ significantly across categories.\\
% \textbf{Follow-up outcomes across chat types.}
% We examined whether four follow-up outcomes: same-file rework, bug/fix follow-up, reverts, and test-touching commits---varied by chat type across multiple post-chat time windows. 
% First, same-file rework differed significantly across chat types only at the 30-day window ($p=.003$, $q=.017$), increasing from 55.8\%--70.3\% within one day to 75.7\%--85.4\% within 30 days. 
% Second, bug/fix follow-up showed the strongest chat-type difference immediately after the chat ($p=.003$, $q=.017$ at one day), but this difference is insignificant at later windows. 
% In contrast, reverts were rare overall (0.0\%--7.2\%), and test-touching commits varied descriptively across chat types (9.9\%--35.0\%); however, neither outcome showed statistically significant differences.
% Overall, chat type differentiated some follow-up behaviors, especially immediate bug/fix activity and longer-horizon rework, but it did not show robust associations with reverts or test-touching commits.\\
\textbf{Chat purpose across project types.}
We next examined whether repositories' chat-purpose distributions or dominant purposes differed by project content category or OSS4SG status. 
Overall, we found no statistically reliable associations. Descriptively, Creativity-Entertainment projects had a higher share of Failure Reporting as the dominant purpose than the overall sample (19.4\% vs. 12.7\%), as did OSS4SG repositories (16.9\% vs. 12.7\%), but neither pattern was significant. 
Thus, project content category and OSS4SG status did not reliably explain variation in dominant chat purpose or chat-purpose distribution.

\par\addvspace{0.5\baselineskip}
\noindent\makebox[\columnwidth][c]{%
\setlength{\fboxsep}{3pt}%
\setlength{\shadowsize}{1pt}%
\shadowbox{%
\parbox{\dimexpr\columnwidth-2\fboxsep-2\fboxrule-\shadowsize\relax}{%
\footnotesize
\setlength{\baselineskip}{9pt}
AI use in OSS was early-concentrated, heavier in smaller and less collaborative repositories,
and centered on Code Writing. 
Chats often preceded activity, but purpose weakly predicted follow-up commits, which were mostly source- or documentation-only, with no robust differences in size, breadth, or project type.}}}

\begin{figure}[t]
  \centering
  \includegraphics[width=0.9\columnwidth]{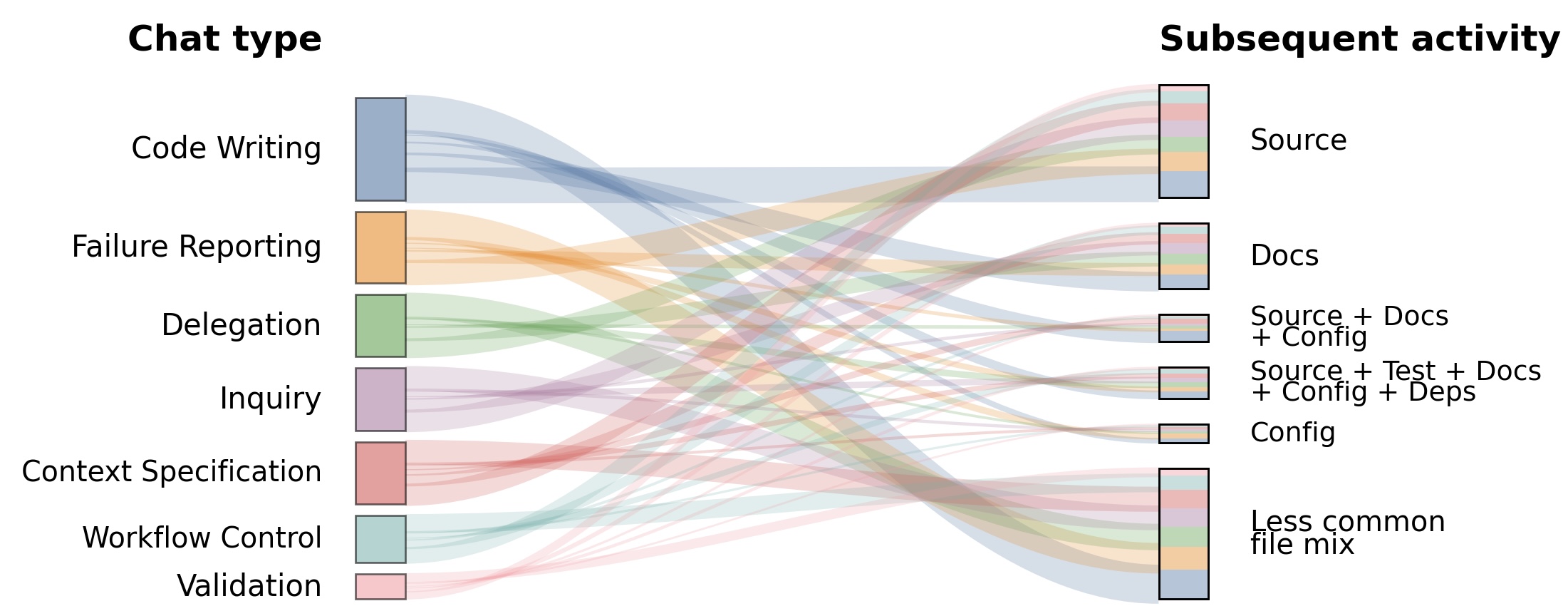}
  \caption{Chronological flow from AI-chat purpose to the file-type composition of the subsequent commit in the same repository. Rare file-type combinations are grouped as ``Less common file mixes.''}
\label{fig:chat-purpose-development-flow}
\end{figure}

\subsection{RQ2: How do OSS repositories' dynamics change after AI adoption?}
% We next compared repositories before and after AI adoption across commit activity, observable code-quality signals, and collaboration patterns, capturing changes in development volume, maintenance and validation signals, and contributor participation.
\subsubsection{How does repository commit activity change}
\label{sec:rq2-1}
Since repositories had unequal pre- and post-adoption observation windows, we compared activity using repository-month rates relative to the adoption
date~\cite{zhao2017impact,ait2022empirical,viggiato2019code,fang2026contribution}, excluding the adoption month to avoid transition-period effects.
Monthly activity declined after adoption: mean commits per repository-month
fell from 19.2 to 7.7, and the median fell from 4.8 to 1.0
($p<.001$, $q<.001$). 
Source-code commits showed the same pattern, with the
mean decreasing from 13.4 to 4.7 and the median from 2.3 to 1.0
($p<.001$, $q<.001$). The validation cohort showed the same direction
of change.

\begin{figure}[t]
    \centering
    \includegraphics[width=0.7\linewidth]{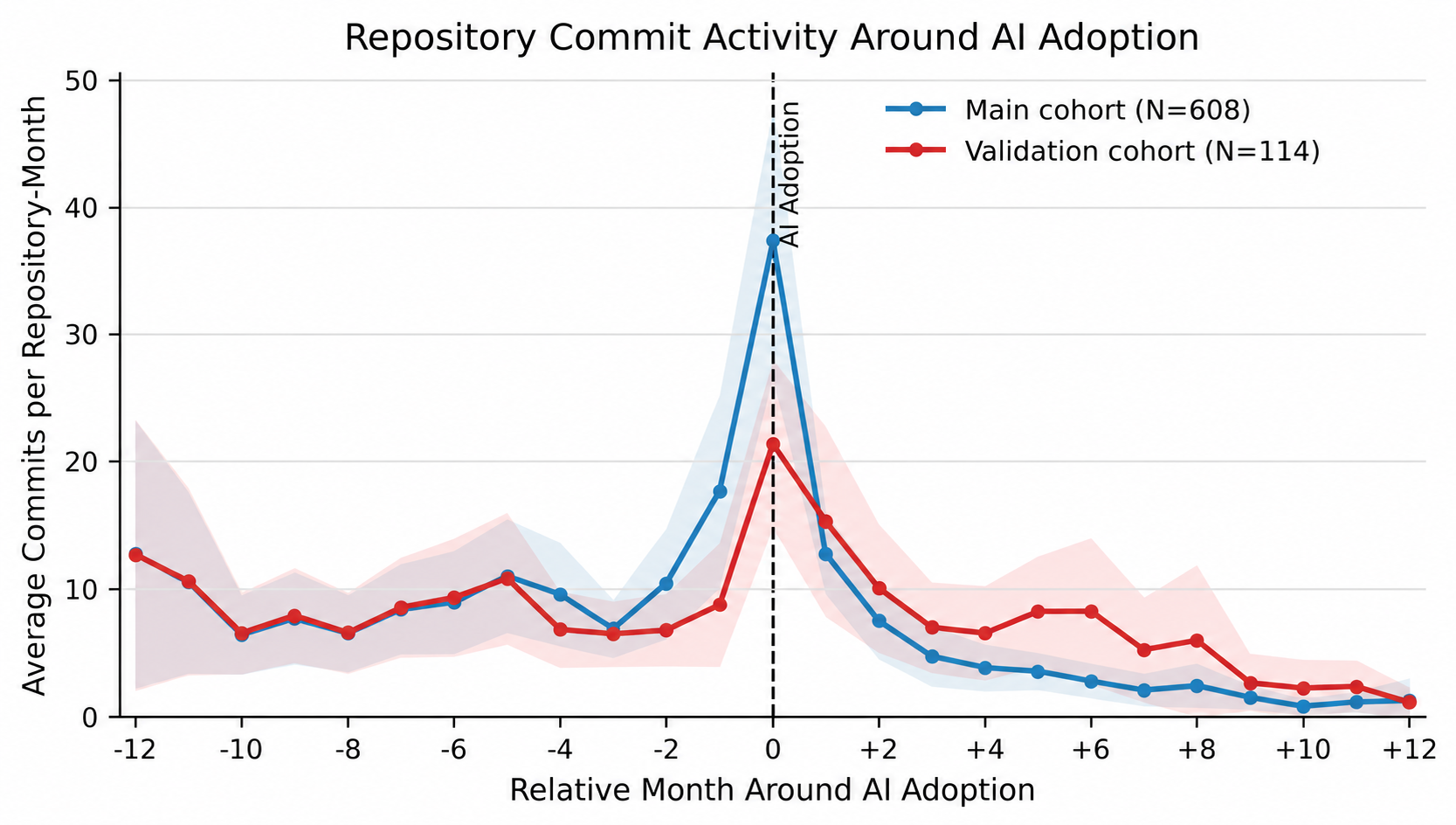}
    \caption{Repository Monthly Commit Activity Relative to AI Adoption.}
    \label{fig:rq2-1}
\end{figure}

We then fit repository-month interrupted time-series (ITS) models with
repository fixed effects, repository age, and repository-clustered standard
errors. In the main cohort, activity was increasing before adoption and
showed no reliable immediate post-adoption jump; the post-adoption trend
turned negative, though this change was not statistically significant for
total commits ($-2.00\%$) or source-code commits ($-1.36\%$). The validation cohort showed the same pattern more clearly, with a short
post-adoption burst followed by a significant decline
(Figure~\ref{fig:rq2-1}; the cohort trends overlap on the left because most sampled repositories are young). 
The two cohorts point in the same direction: a short-lived post-adoption increase followed by a decline rather than sustained growth.
As a sanity check, this pattern matches prior evidence on Cursor
adoption~\cite{he2025speed}, lending external support to the burst-then-decline interpretation.
\subsubsection{Is AI adoption associated with changes in code-quality signals in OSS repositories}
\label{sec:rq2-2}
We examine defect-related activity, testing behavior, and CI outcomes as observable code-quality signals. General changes are summarized in Table~\ref{tab:rq2-summary-signals}.\\
\textbf{Defect-related activity.}
We measured defect-related activity using bug/fix commits and bug-labeled issues, inspired by prior work~\cite{mockus2000identifying,sliwerski2005changes,bachmann2010missing,herzig2013s}. 
We observed a significant rise in bug/fix commit share, while bug-labeled issues showed no meaningful change. 
The mean bug/fix commit share per
repository increased from 9.6\% (pre-adoption) to 16.9\% (post-adoption),
with the median rising from 0.0\% to 11.1\% ($p<.001$, $q<.001$). Raw
bug/fix commit counts did not follow the same pattern: the mean declined
slightly from 17.3 to 16.1, while the median rose from 0.0 to 2.0 ($p<.001$, $q<.001$).
Bug-labeled issues remained largely stable, with means of 0.2 (pre) and
0.1 (post) and medians of 0.0 in both periods. The bug-labeled issue
share rose non-significantly from a mean of 5.5\% to 7.3\%, while the median
stayed at 0.0\% in both periods.
This pattern suggests the rise in bug/fix commit share mainly reflects the drop in overall commit volume (Sec~\ref{sec:rq2-1}), not an actual increase in defect-related activity: the same or fewer bug/fix commits now make up a larger share of a smaller total.

This interpretation is supported by the relative-month interrupted time-series (ITS) models, which showed no robust post-adoption increase in defect-related activity. 
Bug/fix commits rose non-significantly by 13.3\% immediately after adoption and then declined non-significantly month over month. 
Bug/fix commit share rose non-significantly by 3.63\% immediately after adoption, with an essentially flat post-adoption trend. 
Bug-labeled issues and their share likewise showed no reliable immediate change or post-adoption trend.\\
\textbf{Testing and CI.}
For testing and CI, we examined the test-touching commit share and CI outcomes and found no broad change in testing behavior after AI adoption. 
Test-touching commit share rose slightly on average (5.2\% to 6.0\%), with the median unchanged at 0.0\% in both periods, indicating that most repositories did not touch tests more often. 
% Raw test-line changes rose sharply on average (1{,}383 to 3{,}755 lines; $p<.001$,
% $q<.001$), but the median remained 0, suggesting this increase was driven
% by a small subset of repositories rather than a broad shift. 
CI outcomes were essentially stable: mean failure rates changed from 29.1\% to 27.4\% (median 12.2\% to 13.8\%), and mean success rates from 70.9\% to 72.6\% (median 87.8\% to 86.2\%). 
% Thus, the rise in test-line churn did not
% translate into a measurable change in build reliability.

The relative-month ITS models support this interpretation. 
Test-touching commit share showed a positive but non-significant immediate change (+6.7\%). 
% Test lines changed showed a significant immediate jump of
% 55.5\% ($p=.012$, $q=.038$), followed by a significant 4.6\% monthly
% decline ($p=.005$, $q=.023$) --- consistent with a short-lived spike
% rather than a sustained shift in testing practice. 
CI failure and success rates also showed no robust immediate change or post-adoption trend. The validation cohort showed broadly consistent but more pronounced trends.
Thus, we find no clear evidence linking AI adoption to increased testing activity or changed CI outcomes.

Overall, we find no evidence of broad code-quality deterioration after AI adoption. The higher bug/fix share appears driven by reduced commit volume rather than sustained growth in defect-related activity, while testing and CI signals remained stable.

\subsubsection{How do Issues and Pull Requests change}
We also analyzed issue and pull-request activity before and after AI
adoption. Issue activity was sparse: mean issue openings decreased from
3.6 to 0.8 per repository and mean closings from 2.7 to 0.6, with
medians of 0 in both periods; neither the opening nor the closing change reached significance. Issues did, however, make up a smaller share of opened items after adoption, declining from 29.6\% to 19.2\% on average
($p=.001$, $q=.002$), while the issue closure-to-opening ratio increased
from 40.0\% to 95.7\% ($p<.001$, $q=.001$).
PR activity showed a significant before/after shift in raw volume: mean
PR openings increased from 5.1 to 6.3, while the mean merged PRs changed only
slightly from 4.1 to 3.9 (both $p<.001$, $q<.001$), with medians
remaining 0. 
The PR-open share also increased, from 70.4\% to 80.8\% ($p=.001$, $q=.002$), whereas the PR merge share did not decrease significantly.

Despite these before/after differences, relative-month ITS models showed
no reliable immediate or trend-level changes in issue openings, issue
closings, PR openings, or merged PRs. In contrast, issue-resolution time
increased substantially after adoption: the mean rose from 10.5 to 29.4
days and the median from 6.6 to 50.1 days ($p=.002$, $q=.008$). 
PR merge time, by comparison, showed no significant increase.
% (mean: 4.4 to 4.7
% days; median: 0.4 to 0.2 days). 
The validation cohort showed the
same pattern but with weaker statistical support. Results were summarized in Table~\ref{tab:rq2-summary-signals}.

Thus, the main change was compositional: activity became more
PR-oriented and less issue-oriented, with issues growing rarer but
taking substantially longer to resolve once opened. This points to a
reallocation of effort toward PR-based contribution rather than a net
expansion of development activity, alongside reduced responsiveness to
the issues that remain.

\subsubsection{Is AI adoption associated with changes in OSS collaboration patterns}
Contributor participation broadened modestly after AI adoption, but communication and review were still highly concentrated.  
On average, active contributors increased from 1.4 to 1.7 per repository, although the median remained 1.0 in both periods (\(p<.001\), \(q<.001\)). 
The active-contributor share also increased, from a mean of 78.9\% before adoption to 86.8\% after adoption, while the median remained 100\% in both periods (\(p<.001\), \(q<.001\)).  The validation cohort did not show the same pattern: active-contributor share decreased from 87.9\% to 74.1\% on average, with the median remaining 100\% in both periods (\(p=.002\), \(q=.004\)). Thus, AI adoption was not consistently associated with broader contributor participation across cohorts, and the main-cohort pattern may reflect cohort composition rather than a generalizable effect.

Communication intensity changed little: comments per issue/PR increased from 1.4 to 2.0, and PR review comments per PR from 1.7 to 2.1, but neither change was statistically significant; unique commenters and reviewers per repository remained low and stable.
We also examined whether this activity became more or less
concentrated among participants: none of the concentration measures (e.g.,
top-commenter share, commenter HHI, or reviewer HHI) changed significantly, and all remained high throughout. 
The validation cohort showed the same overall lack of broadening in communication and review activity. 
Therefore, the communication and review were already highly concentrated before adoption and remained unchanged afterward. Results were summarized in Table~\ref{tab:rq2-summary-signals}.

\par\addvspace{0.5\baselineskip}
\noindent\makebox[\columnwidth][c]{%
\setlength{\fboxsep}{3pt}%
\setlength{\shadowsize}{1pt}%
\shadowbox{%
\parbox{\dimexpr\columnwidth-2\fboxsep-2\fboxrule-\shadowsize\relax}{%
\footnotesize
\setlength{\baselineskip}{9pt}
AI adoption was associated with short-term activity shifts (e.g., increased bug-fix
commit share ($p<.001$), slower issue resolution ($p=.002$), and more contributor participation ($p<.001$)), but not with sustained productivity gains, code-quality decline, or broader OSS collaboration.
}}}
% We next examined communication intensity, beginning with comments per issue or pull request. 
% We found a non-significant increased from a mean of 1.4 to 2.0 comments per issue/PR, with the median increasing from 0.7 to 1.4. 
% Code-review feedback intensity also showed no significant change: PR review comments per PR changed from a mean of 1.7 to 2.1, while the median remained 0.0 in both periods. Participation in communication and review also remained stable: unique issue/PR commenter changed from a mean of 0.7 to 0.3 per repository, with median 0.0 in both periods, and unique PR reviewers changed from 0.11 to 0.09, again with median 0.00 in both periods.
% Furthermore, we examined whether communication and review activity became more or less concentrated among participants. 
% None of the concentration measures changed significantly. 
% The top-commenter share changed from a mean of 74.4\% to 75.5\%, with the median increasing from 69.7\% to 85.7\%. Commenter HHI changed from a mean of 0.66 to 0.71, with the median increasing from 0.58 to 0.76. 
% Reviewer HHI changed from a mean of 0.81 to 0.77, with the median remaining 1.0 in both periods. 
% These results indicate that communication and review remained concentrated among a small number of participants.
% The older validation cohort showed a similar lack of broadening in communication and review activity.
\newcommand{\inc}{\ensuremath{\nearrow}}
\newcommand{\dec}{\ensuremath{\searrow}}
\newcommand{\siginc}{\ensuremath{\bm{\nearrow}\mkern-3mu^{\bm{*}}}}
\newcommand{\sigdec}{\ensuremath{\bm{\searrow}\mkern-3mu^{\bm{*}}}}
\begin{table}[t]
\centering
\scriptsize
\setlength{\tabcolsep}{4pt}
\renewcommand{\arraystretch}{0.96}
\caption{Summary of repository-signal changes before and after AI adoption. 
Diagonal arrows show the main-cohort direction; bold arrows with * indicate FDR-significant changes. ``Validation'' indicates whether the result showed the same pattern in the validation cohort.}
\label{tab:rq2-summary-signals}
\begin{tabular}{@{}llcc@{}}
\toprule
\textbf{Category} & \textbf{Measure} & \textbf{Main} & \textbf{Validation} \\
\midrule
\multicolumn{4}{@{}l}{\textit{Defect-related activity}} \\
 & Bug/fix commit share        & \siginc & \cmark \\
 & Bug-labeled issue share     & \inc    & \cmark \\
\addlinespace[1pt]

\multicolumn{4}{@{}l}{\textit{Testing and CI}} \\
 & Test-touching commit share  & \inc    & \cmark \\
 & CI outcomes                 & \dec    & \cmark \\
\addlinespace[1pt]

\multicolumn{4}{@{}l}{\textit{Issues and pull requests}} \\
 & Issue opening rate          & \dec    & \cmark \\
 & Issue closing rate          & \dec    & \cmark \\
 & Issue resolution time       & \siginc & \cmark \\
 & PR opening rate             & \siginc & \cmark \\
 & PR merge rate               & \dec    & \cmark \\
 & PR merge time               & \inc    & \cmark \\
\addlinespace[1pt]

\multicolumn{4}{@{}l}{\textit{Collaboration and communication}} \\
 & Active contributors         & \siginc & \xmark \\
 & Contributor concentration   & \sigdec & \xmark \\
 & Communication concentration & \inc    & \cmark \\
\bottomrule
\end{tabular}
\vspace{-6pt}
\end{table}

\subsection{RQ3: How do developers perceive AI tool usage through vibe coding in OSS?}
Respondents generally viewed AI coding assistants as beneficial for OSS
contribution. Most (76\%) agreed that AI assistance lowers barriers to
contribution, and at the construct level, the OSS contribution access
scale was above neutral (mean $= 3.8$, $p=.002$).
More specifically, 80.0\% agreed that AI support makes them more comfortable making a first OSS contribution, and 68\% agreed that it makes them more confident about contributing to projects they had previously hesitated to join.

At the same time, respondents expressed quality and maintenance concerns,
particularly about AI-generated code produced by others. 
Concern about others' AI-generated code was significantly above neutral (mean $= 3.9$, $p=.009$), while concern about one's own AI-generated code was lower and
not statistically significant (mean $= 3.1$). 
Consistent with this asymmetry, a separate paired comparison focused specifically on maintenance burden found that respondents rated others' AI-generated code as more likely to require extra maintenance than their own ($p=.029$).
Respondents also generally believed that human contributors remain
responsible for AI-assisted code, and that OSS communities need clearer
disclosure guidelines (mean $= 4.2$, $p<.001$).

Respondents did not report strong general social-image concerns about
using AI coding assistants in OSS (mean $= 2.9$). However, while 68\% said they would be willing to share their chat history, the remaining respondents cited two main barriers: reputational risks, such as judgment, stigma, or appearing incompetent, and practical risks, such as increased reviewer burden or exposing ideas to competitors or AI vendors. 
These concerns suggest disclosure feels costly to some developers unless it offers a clear benefit to the contributor or community.
Respondents also distinguished appropriate from risky AI use cases in OSS. Most considered AI coding assistants appropriate for
exploring solutions before writing final code (92\%), fixing small bugs
or typos (84\%), improving documentation (84\%), and implementing small, well-scoped features (80\%). 
In contrast, respondents most often felt AI coding assistants should be avoided for security-related changes
(48\%), large or complex feature development (44\%), and
performance-critical changes (36\%).

\par\addvspace{0.5\baselineskip}
\noindent\makebox[\columnwidth][c]{%
\setlength{\fboxsep}{3pt}%
\setlength{\shadowsize}{1pt}%
\shadowbox{%
\parbox{\dimexpr\columnwidth-2\fboxsep-2\fboxrule-\shadowsize\relax}{%
\footnotesize
\setlength{\baselineskip}{9pt}
Developers viewed AI coding assistants as making OSS contributions more accessible but still risky. They were more concerned about others' AI-generated code than their own ($p=.029$), and although most were willing to share chat histories, concerns remained around reputation, reviewer burden, and idea exposure.
}}}
% \vspace{-6pt}

\section{Threats to Validity}
\label{sec:limitation}
\subsubsection{Internal}
\label{sec:limitation-1}
Our sample reflects the long-tailed nature of OSS development: many repositories are small, personal, or low-activity projects, while a few are much larger and more active~\cite{crowston2005social, schweik2013preliminary}. Accordingly, repositories vary substantially in scale, collaboration structure, and development intensity, so we report both means and medians where appropriate. Because most repositories in our sample (90\%) were created in 2025 or later and are relatively small or solo-maintained, generalizability to larger, established projects may be limited. 
To assess robustness, we repeated the main analyses on a validation repository cohort with more established teams and maintenance routines; the main patterns held, suggesting that results are not driven by the newer, smaller-repository majority. As an additional sanity check, our commit-activity trends were consistent with those reported by~\cite{he2025speed}. 
Beyond sample composition, our time-series models control for repository fixed effects and age, but unobserved factors such as developer experience, governance practices, or concurrent workflow changes may still confound the results.
Another concern is causality. Our analyses identify changes following
the first observed AI-chat session, but they do not establish that AI
adoption caused those changes. Repositories may adopt AI while
undergoing other changes (e.g., shifts in contributors, project phase, or maintenance practices) that could independently affect commits, testing, issues, pull requests, and CI outcomes. We therefore interpret our results as associations with AI adoption rather
than causal effects.

\subsubsection{External}
\label{sec:limitation-2}
First, our findings may not extend to private repositories, proprietary codebases, or organizations with development practices that differ from public OSS projects; we therefore limit our claims to public OSS development. Second, our survey was restricted to developers from repositories using Copilot, Cursor, or Claude Code and yielded only 25 valid responses from those with public email addresses who chose to participate, so it may not represent the broader OSS contributor population or the full range of perspectives on AI-assisted development. To partially address this, we triangulate survey responses with the repository-level commit and issue/PR data described above, so our main conclusions do not rely on survey evidence alone. Third, our findings reflect the specific observation period and the AI tools available at that time; as AI coding assistants and vibe coding workflows continue to evolve rapidly, the observed patterns may not generalize to future tool generations or usage patterns.

\subsubsection{Construct}
\label{sec:limitation-3}
Our chat-category and project-type labels were derived through LLM-based classification with manual validation, following prior work~\cite{tang2026programming,fang2023four}. This approach enabled large-scale labeling, but may still introduce errors. 
To mitigate this, we adopted the same labeling procedure and taxonomy validated in prior work~\cite{tang2026programming,fan2023large}; the resulting label distribution broadly resembled that of the original study, and validation against human judgments achieved an F1 score of 0.83 for chat categories and 90.5\% accuracy for project types.
A second concern is that our metrics (e.g., commits, test-file changes, CI pass/fail rates) are proxies for development activity rather than direct measures of correctness, maintainability, or developer intent, and may reflect non-AI-related factors such as bulk renaming or formatting changes. 
We mitigate this by distinguishing source-code commits from broader commit categories, though some measurement imprecision remains.

% \vspace{-4pt}

\section{Discussion}
\label{sec:discussion}
\subsection{From Private Chats to Reviewable Context}
\label{sec:discussion-1}
Our results show that AI chat in OSS is not a single, uniform activity. 
Although Code Writing dominated, developers also used AI for Failure Reporting, Delegation, Inquiry, Context Specification, Validation, and Workflow Control. 
These purposes showed distinct downstream patterns, most often leading to source-code commits and, secondarily, documentation-only edits; Workflow Control and Validation were more often associated with broader multi-file changes.
This suggests that AI chat is not merely a code-generation channel but also supports debugging, explanation, planning, validation, and coordination.

At the same time, the dominant chat purpose did not vary significantly by project type or OSS4SG status. This suggests that developers use AI chat in broadly similar ways across project domains. The more important OSS-specific issue may be what happens after these conversations: although nearly all chat sessions were followed by commits, the purpose and reasoning behind the AI interaction are usually not visible in the repository. A commit may show the final code change, but not whether AI was used to debug a failure, specify project context, validate a solution, or plan a broader update.

This creates an opportunity for lightweight transparency mechanisms in OSS platforms and workflows. For example, developers could summarize relevant AI assistance in commit messages or pull-request descriptions, indicating whether AI was used for implementation, debugging, validation, or planning. Since 68\% of surveyed developers indicated willingness to share chat history, some contributors may be open to surfacing this context, but full transcript sharing should not be required. Instead, structured summaries or ``AI-assisted'' tags could provide useful review context while reducing concerns about stigma, reviewer burden, or exposure of sensitive ideas. Maintainers could then calibrate review effort based on the risk and scope of the change, rather than treating all AI-assisted contributions the same.

\subsection{From Individual Adoption to Ecosystem-Level Consequences}
\label{sec:discussion-2}
% Our results suggest that AI coding assistants are heavily used in smaller, more solo-oriented repositories, where AI accounts for a larger share of activity than in larger, older projects. This matters because 
% AI adoption does not necessarily imply community-wide adoption: in many repositories, AI use appears to be driven by the primary maintainer or a small group of active developers. 
% Even 
Even after AI adoption, communication and review remained concentrated among a few contributors, despite survey responses suggesting that AI lowers barriers to OSS contribution. 
This suggests that AI may help contributors work more independently by supporting code understanding, debugging, and implementation, but it may not necessarily make collaboration more visible or distributed. 
However, if problem-solving increasingly occurs in private AI chat sessions, important design reasoning and debugging context may become less accessible to maintainers and future contributors. Rather than showing that team-based programming becomes less important, these findings raise a broader question: whether AI-assisted OSS contribution shifts collaboration from shared problem solving toward post-hoc review, documentation, and accountability.

Future work should therefore examine both AI adopters and non-adopters to better understand their concerns, barriers, and expectations, as well as how AI adoption evolves across OSS communities over time. Such evidence could inform OSS contribution guidelines that support open communication and collaboration, and AI tool designs that support not only code generation but also effective communication.
Future research could also investigate how AI affects developer behavior, review expectations, contribution transparency, and project maintainability. Chat data offers a valuable window into developer intent, but should be carefully linked to repository outcomes such as commits, tests, issues, and pull requests to draw meaningful conclusions. Researchers should further distinguish between AI use in newly bootstrapped projects and AI adoption in mature, collaborative projects, as these contexts likely involve different risks, benefits, and governance needs.

Our findings also raise platform-level questions. If AI-assisted development increasingly produces small or solo-maintained repositories, platforms such as GitHub may need to better support how AI-heavy projects are created, reviewed, maintained, and sustained, especially as AI-driven activity may strain existing infrastructure~\cite{stewart2026githubcapacity}. One option is to surface AI-tool usage as repository metadata, similar to language or license information, through maintainer self-reporting, tool integrations, or inferred activity patterns. Such metadata could help contributors assess a project’s maintenance model before contributing to or depending on it. Community practices are equally important: although some tools already mark AI involvement through co-author trailers, standardized lightweight disclosure remains limited~\cite{moraes2026githubdiscussion,register2026githubslop}. Maintainers could also require targeted review steps for AI-heavy contributions, such as tests or human-written documentation, to keep design reasoning visible when initial problem-solving occurs in private AI sessions.
% \vspace{-5pt}

\section{Conclusion}
\label{sec:conclusion}
AI coding assistants have become conversational collaborators that support \textit{vibe-coding}, where developers guide AI-generated code through natural language. Yet little is known about how these interactions relate to subsequent OSS activity. We collected 13{,}360 AI conversation sessions across 1{,}356 OSS repositories, linked them to repository histories, and complemented the results with a developer survey.
We found AI use concentrates in smaller, less mature, less collaborative repos. Post-adoption, projects generally gained contributors and reduced contributor concentration, though communication stayed concentrated. 
Code Writing dominated chat use, most sessions led to commits, and code quality/merge rates held steady. Surveys show developers see AI as lowering contribution barriers, but worry about others' AI-generated code, reviewer burden, perceived incompetence, and idea exposure. These offer a large-scale empirical view of AI-assisted OSS contribution and inform responsible vibe-coding governance.
\newpage
\bibliographystyle{IEEEtran}
\bibliography{my}
\end{document}